\documentclass{article}
\usepackage{graphicx} 
\usepackage{bm}
\usepackage{verbatim, color, array}

\usepackage{amsthm, amssymb}
\usepackage{float}
\usepackage{amsmath}
\usepackage{hhline}
\usepackage{booktabs}
\usepackage{soul}
\usepackage[super]{natbib}
\usepackage[table]{xcolor} 
\usepackage{hyperref}
\usepackage{multirow}
\usepackage{nicematrix}
\usepackage{threeparttable}
\usepackage{enumitem}
\usepackage{tikz}
\usepackage{appendix}
\usetikzlibrary{shapes, arrows, positioning, calc, shadows, fit}
\usepackage{longtable}
\usepackage[affil-it]{authblk} 

\title{\textbf{PKBOIN-12: A Bayesian optimal interval Phase I/II design incorporating pharmacokinetics outcomes to find the optimal biological dose}}
\author[1]{Hao Sun \thanks{Corresponding author: *Hao Sun, Global Biometrics \& Data Sciences, Bristol Myers Squibb, New Jersey, USA. Email: \href{mailto:hao.sun@bms.com}{hao.sun@bms.com}}}
\author[2]{Jieqi Tu}
\affil[1]{Global Biometrics \& Data Sciences, Bristol Myers Squibb, New Jersey, USA}
\affil[2]{Division of Epidemiology and Biostatistics, School of Public Health, University of Illinois at Chicago, Illinois, USA}

\date{December 2023}

\begin{document}

\maketitle
\begin{abstract}
Immunotherapies and targeted therapies have gained popularity due to their promising therapeutic effects across multiple treatment areas. The focus of early phase dose-finding clinical trials has shifted from finding the maximum tolerated dose (MTD) to identifying the optimal biological dose (OBD), which aims to balance the toxicity and efficacy outcomes, thereby optimizing the risk-benefit trade-off. These trials often collect multiple pharmacokinetics (PK) outcomes to assess drug exposure, which has shown correlations with toxicity and efficacy outcomes but has not been utilized in the current dose-finding designs for OBD selection. Moreover, PK outcomes are usually available within days after initial treatment, much faster than toxicity and efficacy outcomes. To bridge this gap, we introduce the innovative model-assisted PKBOIN-12 design, which enhances the BOIN12 design by integrating PK information into both the dose-finding algorithm and the final OBD determination process. We further extend PKBOIN-12 to the TITE-PKBOIN-12 design to address the challenges of late-onset toxicity and efficacy outcomes. Simulation results demonstrate that the PKBOIN-12 design more effectively identifies the OBD and allocates a greater number of patients to it than BOIN12. Additionally, PKBOIN-12 decreases the probability of selecting inefficacious doses as the OBD by excluding those with low drug exposure. Comprehensive simulation studies and sensitivity analysis confirm the robustness of both PKBOIN-12 and TITE-PKBOIN-12 designs in various scenarios. 
\end{abstract}


\section{Introduction}
The 2000s marked the rise of targeted therapies, such as imatinib (Gleevec) and trastuzumab (Herceptin), ipilimumab (Yervoy), and pembrolizumab (Keytruda). These medications specifically target and destroy cancer cells by locking onto unique molecular alterations predominantly found in those cells. An increasing number of targeted therapies are now standard treatments for many cancers. For example, immune checkpoint inhibitors have gained widespread application in the treatment of various cancer types, including melanoma, lung cancer, kidney cancer, bladder cancer, and lymphomas. With the trend of current oncology drug development shifting from traditional chemotherapies to immunotherapies and molecularly targeted therapies, the objective of early phase clinical trials also shifted from determining the maximum tolerated dose (MTD) to identifying the optimal biological dose (OBD), which aims to optimize the toxicity-efficacy trade-off by accounting for both toxicity and efficacy outcomes. For instance, the chimeric antigen receptor (CAR) T therapies require a balance of boosting the immune system to battle against cancer while avoiding over-stimulation. Unlike traditional cytotoxic agents in chemotherapies, where efficacy and toxicity are generally assumed to increase with dose, novel agents for targeted therapies may not exhibit a monotonic pattern in the dose-response relationship. Conventional clinical dose-finding trial designs using parametric approaches become more difficult since there are no constraints on the dose-response curve.

The current existing dose-finding designs can be categorized into rule-based, model-assisted, and model-based designs. The 3+3 design is the most commonly used rule-based design because of its simplicity and ease of implementation \cite{araujo2021contemporary}. Commonly used model-based dose-finding designs for MTD estimation include the CRM design \cite{o1990continual}, the EWOC design \cite{Babb1998}, the BLRM design \cite{neuenschwander2008critical}, and EWOC-NETS design which is an extension of the EWOC design to improve trial efficiency and reduce loss of information \citep{Chen2012}. Model-based dose-finding designs aimed at OBD detection include the Eff-Tox design \cite{thall2004dose}, the CP-logistic design \cite{sato2016adaptive}, the L-logistic design \cite{zang2014adaptive}, and the B-dynamic design \cite{liu2016robust}. On the other hand, some popular model-assisted dose-finding designs include the mTPI design\cite{ji2010modified}, the mTPI-2 design\cite{li2017toxicity}, and the BOIN design \cite{liu2015bayesian, yuan2016bayesian}. To optimize the toxicity-efficacy trade-off, an increasing number of model-assisted designs have been developed, such as the BOIN-ET design \cite{takeda2018boin}, the BOIN12 design\cite{lin2020boin12}, the STEIN design \cite{lin2017stein}, the TEPI design \cite{li2017toxicity}, the TEPI-2 design \cite{li2020tepi}, the PRINTE design\cite{lin2021probability}, and the uTPI design\cite{shi2021utpi}. Among all these designs, BOIN-12 is the only design that incorporates both toxicity and efficacy outcomes to determine the OBD and has been approved by FDA to be used in early phase clinical trials. For instance, the BOIN12 design was employed in the development of a Phase I/II study of enhanced CD33 CAR-T Cells in subjects with relapsed or refractory acute myeloid leukemia (NCT04835519) and a Phase I trial which aimed to evaluate the safety and tolerability of CD5 CAR-T cells in subjects with relapsed or refractory T-cell acute lymphoblastic leukemia (NCT05032599).

Rapid patient enrollment and delayed onset of toxicity and efficacy outcomes present additional challenges for certain non-cytotoxic anti-cancer agents, as these factors can slow the assessment process. When the accrual rate is high, some participants may not have finished their outcome evaluations by the time interim decisions need to be made, leading to delays in dose allocation for incoming patient cohorts. Similar issues can occur when toxicity and/or efficacy have a delayed onset, necessitating a longer observation period to accurately assess outcomes. In order to shorten the trial duration, many time-to-event (TITE) designs have been developed based on either model-based designs or model-assisted designs, such as TITE-CRM \cite {cheung2000sequential}, TITE-BOIN \cite{yuan2018time}, TITE-keyboard \cite{lin2020time}, TITE-EffTox \cite{jin2014using}, TITE-BOIN12 \cite{zhou2022tite}, and TITE-BOIN-ET \cite{takeda2020tite} designs. Based on extensive simulations in these studies, these TITE extensions of the dose-finding designs are able to make real-time decisions with pending outcomes. They can not only reduce the trial duration but also sustain a comparable performance on MTD or OBD selection accuracy and patient allocation compared with the original dose-finding designs. 


Another important objective in early phase dose-finding clinical trials is to evaluate the pharmacokinetics (PK) exposure of the target drug. PK outcomes in clinical trials are measurements that describe the disposition of a drug within the body over a period of time, focusing on the processes of absorption, distribution, metabolism, and excretion. For example, area under the curve (AUC), which measures the total exposure to the drug over time, is a vital PK outcome, because it helps to understand the dose-response relationship. Maximum concentration (Cmax), as another important PK outcome, implies the peak plasma concentration of a drug after administration. PK information can be used to assess drug intervention in humans. Exposure-safety and exposure-efficacy analyses frequently form part of the objectives in early phase clinical trial designs. The PK outcomes can be thought of as a good indicator in evaluating the several efficacy outcomes of the investigational agent such as overall response, immune response, and overall survival \cite{Centanni2019, takeda2023bayesian}. Moreover, some evidence also showed that subjects with dose-limiting toxicities (DLT) have a statistically significantly larger AUC value compared with subjects with no severe toxicities \cite{Broker2006}. This suggests that incorporating PK outcomes into early phase dose-finding clinical trials could enhance the accuracy of OBD selection. However, in many such trials, the dose-finding process and PK analyses are often conducted independently \cite{ursino2017}.

There are some dose-finding trial designs in the early phases that have incorporated PK evaluations. Most of these designs are model-based trial designs. Piantadosi and Liu \cite{Piantadosi1996} enhanced the CRM design by integrating PK data as a covariate into a parametric dose-response logit model (PKCOV). Patterson et al. \cite{Patterson1999} and Whitehead et al. \cite{whitehead2001easy} used a hierarchical PK-toxicity model in a cross-over trial setting. Takeda et al. extended the CRM design by incorporating an adjustment based on the latest AUC assessment instead of the standardized dose \cite{takeda2018bayesian}. These designs predominantly centered on pinpointing the MTD based strictly on toxicity outcomes and did not include efficacy outcomes to select the optimal dose considering the benefit-risk balance. Furthermore, these model-based designs incorporating PK assessments may be intricate for researchers lacking advanced statistical expertise. In addition, these designs are not suitable for clinical trials with late-onset outcomes. Based on our knowledge the PK outcomes have not been used in an early phase trial design for OBD selection, especially in the context of late-onset efficacy and toxicity results. Takeda et al. \cite{takeda2023bayesian} proposed a model-assisted design that made use of all toxicity, efficacy, and PK information. However, their design focused on dose optimization and still does not tackle the challenge of late-onset outcomes. 


In this article, we propose a model-assisted PKBOIN-12 design, which is a Bayesian optimal interval design that integrates the toxicity, efficacy, and PK outcomes to select the optimal dose. PKBOIN-12 is an extension of the BOIN12 design that utilizes PK data to improve the performance of OBD selection. As a model-assisted design, PKBOIN-12 does not require complex model fitting in real time during the trial. Additionally, it can directly employ the rank-based desirability score (RDS) table from the BOIN12 framework \cite{lin2020boin12}. Moreover, BOIN12 has already been utilized in several clinical trials. These benefits render PKBOIN-12 more straightforward to implement in actual clinical trials than other designs that integrate PK data. Our design takes into account the correlation between the PK outcomes with both efficacy and toxicity outcomes. It is logical to assume that at lower dose levels, certain PK outcomes, like AUC, would be lower, which in turn might be associated with low efficacy rates. Therefore, we can utilize PK outcomes to eliminate the futile dose levels during the trial and in the final OBD selection procedure after the end of the trial. The traditional BOIN12 design only eliminates an ineffective dose level in the trial if there is an extremely high posterior probability that its efficacy rate is below the predefined minimum efficacy rate. This limitation means that BOIN12 may inadvertently have an innegligible chance of choosing a dose as the OBD that is actually ineffective. By incorporating PK data, our PKBOIN-12 design is more adept at ruling out such ineffective dose levels when determining the OBD, thereby increasing the probability of identifying the true OBD and ensuring more patients are assigned to it compared to BOIN12. For the specific challenge posed by the late-onset efficacy and toxicity outcomes, we have also developed a time-to-event version of this design, the TITE-PKBOIN-12 design. A key benefit of including PK outcomes is the rapidity of PK assessments, which are completed in a few days – much faster than determining efficacy and toxicity results. As a result, with the TITE-PKBOIN-12 design, it is necessary to impute only the probabilities for toxicity and efficacy.


The remainder of this manuscript is organized as follows. In Section~\ref{sec:methodology}, we demonstrate the methodology and design algorithms of our proposed PKBOIN-12 and TITE-PKBOIN-12 designs. Section~\ref{sec:simulation} presents comprehensive simulation studies and sensitivity analyses to evaluate the operating characteristics of our proposed designs. In Section~\ref{sec:software}, we introduce the software developed to implement our proposed designs for practical users. We close with a brief conclusion in Section~\ref{sec:conclusion}. 

\section{Methodology}\label{sec:methodology}
Consider an early phase clinical trial with a total of $D$ doses. The trial collects a binary toxicity outcome $Y_T$ and a binary efficacy outcome $Y_E$, where $Y_T = 1$ indicates a DLT and $Y_E = 1$ represents an efficacy response during the trial. Define $p_T$ and $q_E$ as the maximum acceptable toxicity probability and the minimum acceptable efficacy probability, respectively. The true probabilities of toxicity and efficacy for the dose level $d$ are defined as $p_d$ and $q_d$ respectively, $d = 1, \ldots, D$. $p_d$ is assumed to increase strictly with the dose level, i.e. $p_1 < \ldots < p_D$. This assumption might not hold for the efficacy outcomes, especially under some newly emerging therapies, such as immuno-oncology therapies and molecularly targeted therapies. Let dose level $d_{MTD}$ be the MTD, i.e. $d_{MTD} = \arg\max_{d} p_d I(p_d \leq p_T)$. If all dose levels have $p_d > p_T$, the MTD does not exist and the trial should be early terminated. 

Let $n_{d,T}$ represent the number of patients who experience a DLT and $n_{d,E}$ denote the number of patients who show an efficacy response at dose level $d$. Therefore, the outcomes based on toxicity and efficacy can be categorized into four types: 
$O_1$= ($Y_T = 0, Y_E = 1$); $O_2$ = ($Y_T = 0, Y_E = 0$); $O_3$ = ($Y_T = 1, Y_E = 1$); $O_4$ = ($Y_T = 1, Y_E = 0$). The patient count for each category $O_i$ at dose level $d$ is denoted by $n_{d,i}$ for $i = 1, 2, 3, 4$, satisfying $n_{d,T} = n_{d,3} + n_{d,4}$ and $n_{d,E} = n_{d, 1} + n_{d, 3}$. Let $n_d = \sum_i^4 n_{d, i}$ be the total number of patients assigned to dose level $d$. We further define $\pi_{d,i}$ as the true probability and $\hat{\pi}_{d, i} = n_{d,i}/n_d$ as the observed probability for each category $O_i$. Given the definitions, it follows that $p_d = \pi_{d, 3} + \pi_{d, 4}$, $q_d = \pi_{d, 1} + \pi_{d, 3}$, $\hat{p}_d= n_{d,T}/n_d$, and $\hat{q}_d = n_{d,E}/n_d$.

\subsection{Original BOIN12 Design}\label{sec:BOIN12}
BOIN12 \cite{lin2020boin12} was developed to find the optimal biological dose (OBD) which optimizes the risk-benefit trade-off of the treatment due to the complicated dose-response relationship \cite{guo2017bayesian, liu2018bayesian, yuan2017bayesian}. This design utilizes the utility score value to quantify the risk-benefit trade-off. Let $u_i$ be the utility score of $O_i$ for $i = 1, 2, 3, 4$, and $\bm{u} = (u_1, u_2, u_3, u_4)$ denote the utility vector. We assign utility scores to the outcomes, setting the best outcome $O_1$ at $u_1 = 100$ and the worst outcome $O_4$ at $u_4 = 0$. The utility scores for the remaining two outcomes, $u_2$ and $u_3$, fall within the range of $[0, 100]$ and are typically selected by physicians. Given the utility vector $\bm{u}$ and the outcome probability vector $\bm{\pi}_d = (\pi_{d,1}, \pi_{d,2}, \pi_{d,3}, \pi_{d,4})$, the expected utility $EU_d$ of dose level $d$ is $EU_d = \sum_{i = 1}^4 u_i\pi_{d, i}$. A higher value of $EU_d$ indicates a higher desirability of a dose in the risk-benefit trade-off \cite{lin2020boin12}. In BOIN12, the OBD is the dose level with the highest $EU_d$. When $u_2 + u_3 = 100$, we have $EU_d = u_2(1-p_d) + u_3q_d$. Moreover, if $u_2 = 0$ and $u_3 = 100$, then $EU_d = 100q_d$. The default values of $(u_2, u_3)$ are (40, 60) to optimize the risk-benefit trade-off. However, the summation of $u_2$ and $u_3$ is not necessary to be 100.

BOIN12 models the dose utilities using a quasi-beta-binomial model. Lin et al\cite{lin2020boin12} introduced a rank-based desirability score (RDS) for the dosing decision during the trial. Suppose the standardized desirability of the dose level $d$ as $u_d = EU_d/100$. BOIN12 defines the ``quasi-binomial" event $x_d$ given by $x_d = 100^{-1}\sum_{i=1}^4 u_in_{d,i}$, which can be interpreted as the number of events observed from $n_d$ patients with a quasi-binomial distribution, i.e. $x_d \sim \text{quasi-binomial}(n_d, u_d)$. Under the Bayesian framework, BOIN12 assumes that $u_d$ follows a non-informative uniform prior, specifically $\text{Beta}(1, 1)$, which leads to the posterior distribution of $u_d$ as 
$$
u_d \mid n_d, x_d \sim \text{Beta}(1 + x_d, 1 + n_d - x_d).
$$
RDS is the rank of the posterior probabilities of $\Pr(u_d > u_b\mid n_d, x_d)$ for all possible situations, where the utility benchmark $u_b = \underline{u} + (100 - \underline{u})/2$ with $\underline{u} = u_1 (1-p_T)q_E + u_2 (1-p_T)(1-q_E) + u_3p_Tq_E$. The dose with a higher RDS is equivalent to having a higher posterior probability $\Pr(u_d > u_b\mid n_d, x_d)$. We presented an example of RDS table given $p_T = 0.35$ and $q_E = 0.25$ as the Table S1 in the supplementary document. 

Let $c = 1$ denote the initial cohort and $d$ as the lowest or a pre-specified starting dose level. Define $N^* = 6$ as the pre-specified sample size cutoff. The dose-finding algorithm of BOIN12 follows from Lin et al\cite{lin2020boin12}:
\begin{itemize}
    \item[1. ] Treat the cohort $c$ at the dose level $d$. 
    \item [2. ] Calculate the observed toxicity rate $\hat{p}_d = n_{d,T}/n_d$ at the current dose level $d$ and compare it with two constants $0 < \lambda_1 < \lambda_2 <1$, where $\lambda_1$ and $\lambda_2$ are the two pre-specified parameters depending on $p_T$ from the BOIN design \cite{liu2015bayesian}. 
        \begin{itemize}
            \item[(a) ] if $\hat{p}_{d} \geq \lambda_2$, set $d'$ as the next lower dose level $d - 1$; 
            \item[(b) ] if $\hat{p}_d < \lambda_2$, $n_d \geq 9$, and the next higher dose level $d+1$ has not been used, set $d'$ to the next higher dose level $d+1$; 
            \item[(c) ] if $\lambda_1 < \hat{p}_d < \lambda_2$: (i) when $n_d \geq N^*$, select one dose $d'$ from the admissible set $\{d-1, d\}$ with larger RDS; (ii) when $n_d < N^*$, select one dose $d'$ from the admissible set $\{d-1, d, d+1\}$ with the largest RDS; 
            \item[(d) ] if $\hat{p}_d \leq \lambda_1$, select one dose $d'$ from the admissible set $\{d-1, d, d+1\}$ with the largest RDS.
        \end{itemize}
    \item[3. ] Set $c = c+1$ and update $d = d'$ based on the dosing decision from Step 2. 
    \item[4. ] Repeat Steps 1 - 3 until the maximum sample size is reached. 
\end{itemize}

BOIN12 selects the OBD after the maximum sample size is reached utilizing the following two steps: (1) determine the MTD by applying the isotonic regression to the observed toxicity rates $\{\hat{p}_d\}_{d=1}^D$ to obtain isotonic estimates $\{\tilde{p}_d\}_{d=1}^D$ and choose the dose level $d_{MTD}^*$ as the MTD such that $d_{MTD}^* = \arg\min_d |\tilde{p}_d - p_T|$; (2) select one dose from $\{1, \ldots, d_{MTD}\}$ as OBD with the highest estimated utility. To avoid allocating patients to toxic or ineffective doses, BOIN12 considers the following two elimination criteria:
\newlength{\mylength}
\settowidth{\mylength}{\textbf{(Efficacy)}}
\begin{itemize}[left=0em]
    \item[] \makebox[\mylength][l]{\textbf{(Safety)}}   \parbox[t]{0.9\linewidth}{if $\Pr(p_d > p_T\mid \hat{p}_d, n_d) > C_T$, eliminate dose level $d$ and all above doses from the dose list;}
    \item[] \makebox[\mylength][l]{\textbf{(Efficacy)}} \parbox[t]{0.9\linewidth}{if $\Pr(q_d < q_E\mid \hat{q}_d, n_d) > C_E$, eliminate dose level $d$ from the dose list;}
\end{itemize}
$C_T = 0.95$ and $C_E = 0.9$ are two pre-specified probability cutoffs for toxicity and efficacy, respectively. BOIN12 assumes that $p_d$ and $q_d$ follow non-informative beta priors, $\text{Beta}(1,1)$. During the trial, only the doses fitting the two criteria can be assigned to the next cohort of patients. The trial should be terminated without selecting any dose if all dose levels are eliminated.

\subsection{Proposed PKBOIN-12 Design}\label{sec:PKBOIN_12}
\subsubsection{Assumptions of PK outcome}
BOIN12 only utilizes toxicity and efficacy information for selecting the OBD. However, early phase clinical trials also collect PK outcomes. These important PK components provide information on how the drug is absorbed, distributed, metabolized, and eliminated by the body, which is critical for determining the appropriate dosing regimen for the drug and for predicting its efficacy and safety in later stages of development. Utilizing these PK outcomes can benefit clinical researchers in identifying the OBD, considering that an early phase trial usually has a small sample size with limited information on toxicity and efficacy. Therefore, we propose the PKBOIN-12 design to make use of the PK information to solve the OBD identification issue. 

There are many continuous PK outcomes, such as AUC, Cmax, Tmax, half-life time ($t_{1/2}$), clearance (CL), the volume of distribution (Vd), etc. Therefore, we suppose $r_d$ as the true mean of the targeted continuous PK outcome at the dose level $d$ for analysis, which is presumed to have a connection with both the outcomes of toxicity and efficacy \cite{centanni2019clinical}. Similar to the toxicity outcome, the true PK values $\{r_d\}_{d = 1}^D$ are assumed to increase strictly with the dose level, i.e. $r_1 < \ldots < r_D$. We define $r_P$ as the target continuous PK value, such that a dose level with a low PK outcome, i.e. $r_d < r_P$, is considered to be an ineffective dose. The value of $r_P$ should be carefully specified because the distribution of the PK outcome depends on various factors such as the compound, dosage, administration method, the treated population, etc. An appropriate choice of $r_P$ should ensure that an effective dose level $d$ has $r_d \geq r_P$. However, an overly high PK outcome does not guarantee the efficacy of the dose because of the unknown dose-response relationship. Doses can also be potentially toxic if they have overly high PK outcomes. 

Under the Bayesian framework, we assume that each continuous $r_d$ follows a truncated normal prior, $\text{truncated-N}(0, \sigma_0^2)$ to guarantee $r_d$ is positive, where $\sigma_0$ is assumed to be large. By default, we choose $\sigma_0 = 10000$. Let $I_d$ denote the set of all patients assigned to dose level $d$. We assume that the individual-level PK value $r_{d,j} \stackrel{i.i.d}{\sim} N(r_d, \sigma_d^2)$. Therefore, the posterior distribution of $r_d$ is given by 
$$
r_d \mid \hat{r}_d, \sigma_0^2 \sim \text{truncated-N}(\frac{n_d\hat{r}_d}{\sigma_d^2(1/\sigma_0^2 + n_d/\sigma_d^2)}, \frac{1}{1/\sigma_0^2 + n_d/\sigma_d^2}),
$$
where $\hat{r}_d = n_d^{-1}\sum_{j \in I_d} r_{d,j}$ denotes the observed PK sample mean. Therefore the MAP estimate of $r_d$ can be approximated by $\hat{r}_d$ if the small term $1/\sigma_0^2$ is ignored. 

Let $\zeta_1$ denote the cutoff on the continuous PK outcome, satisfying $0 \leq \zeta_1 \leq r_P$. We refer to Takeda et al \cite{takeda2023bayesian} for deciding $\zeta_1$ by considering two simple hypotheses: 
$$
H_{P0}: r_d = r_P \quad vs \quad H_{P1}: r_d = r_I,
$$
where $r_I$ denotes an inefficacious PK value. By default, we suggest $r_I = 0.6 r_P$, which is also recommended by Mu et al\cite{mu2019gboin} and Takeda et al \cite{takeda2023bayesian}. An appropriate $\zeta_1$ should minimize the posterior probability of incorrect marginal decisions for PK given by 
$$
\Pr(\text{Incorrect PK decisions} \mid r_d) = \pi_{P0}\Pr(\hat{r}_d \leq \zeta_1 \mid H_{P0}) + \pi_{P1}\Pr(\hat{r}_d > \zeta_1 \mid H_{P1}), 
$$
where $\pi_{P0}$ and $\pi_{P1}$ are the prior probabilities of the two hypotheses of the PK value. Under the assumption of equal non-informative prior probabilities for both hypotheses, set at $\pi_{P0} = \pi_{P1} = 1/2$. Therefore, $\Pr(\text{Incorrect PK decisions} \mid r_d) = [\Phi(\zeta_1; r_P, \sigma_d^2/n_d) + 1 - \Phi(\zeta_1; r_I, \sigma_d^2/n_d)]/2$, so that the optimal cutoff value is established as $\zeta_1 = (r_P + r_I)/2 = 0.8r_P$.

\subsubsection{Dose-finding Algorithm}
The dose allocation algorithm of PKBOIN-12 is based on that of the BOIN12 design. Define $d_{PK,min}$ as the lowest dose level such that its observed PK sample mean is greater than the cutoff point, i.e. 
$$
d_{PK, min} = \arg\min_d \hat{r}_d > \zeta_1.
$$
If no dose level achieves $\hat{r}_d > \zeta_1$, then $d_{PK, min}$ does not exist. In PKBOIN-12, we augment the admissible set to include lower dose levels than the current one, provided they have acceptable PK outcomes and less toxicity while maintaining adequate drug exposure. Our proposed design also utilizes the number of quasi-events $x_d$ and the desirability score for the dosing decisions. PKBOIN-12 inherits all the benefits of using utility scores as seen in BOIN12 and achieves greater flexibility in selecting OBD by integrating PK information. Unlike BOIN12, which limits the admissible set to a maximum of three dose levels, our PKBOIN-12 design expands this to encompass more dose levels with efficacious PK outcomes. For simplicity, we define $d^*$ as the minimum dose level between dose level $d-1$ and $d_{PK, min}$, i.e., $d^* = \min\{d-1, d_{PK, min}\}$, such that the new constructed admissible set, $A$, is a subset of $\{d^*, \ldots, d+1\}$. By definition, if $d_{PK, min}$ does not exist, $d^* = d-1$. However, if $d_{PK, min}$ exists and is lower than $d-1$, we will expand the admissible set by adding $\{d^*, \ldots, d-2\}$ to diversify the selection of appropriate dose levels with an efficacious PK value and optimize the desirability for the next cohort of patients. 

\begin{figure}[htbp]
    \hspace{-2cm}
    \includegraphics[width = 8in]{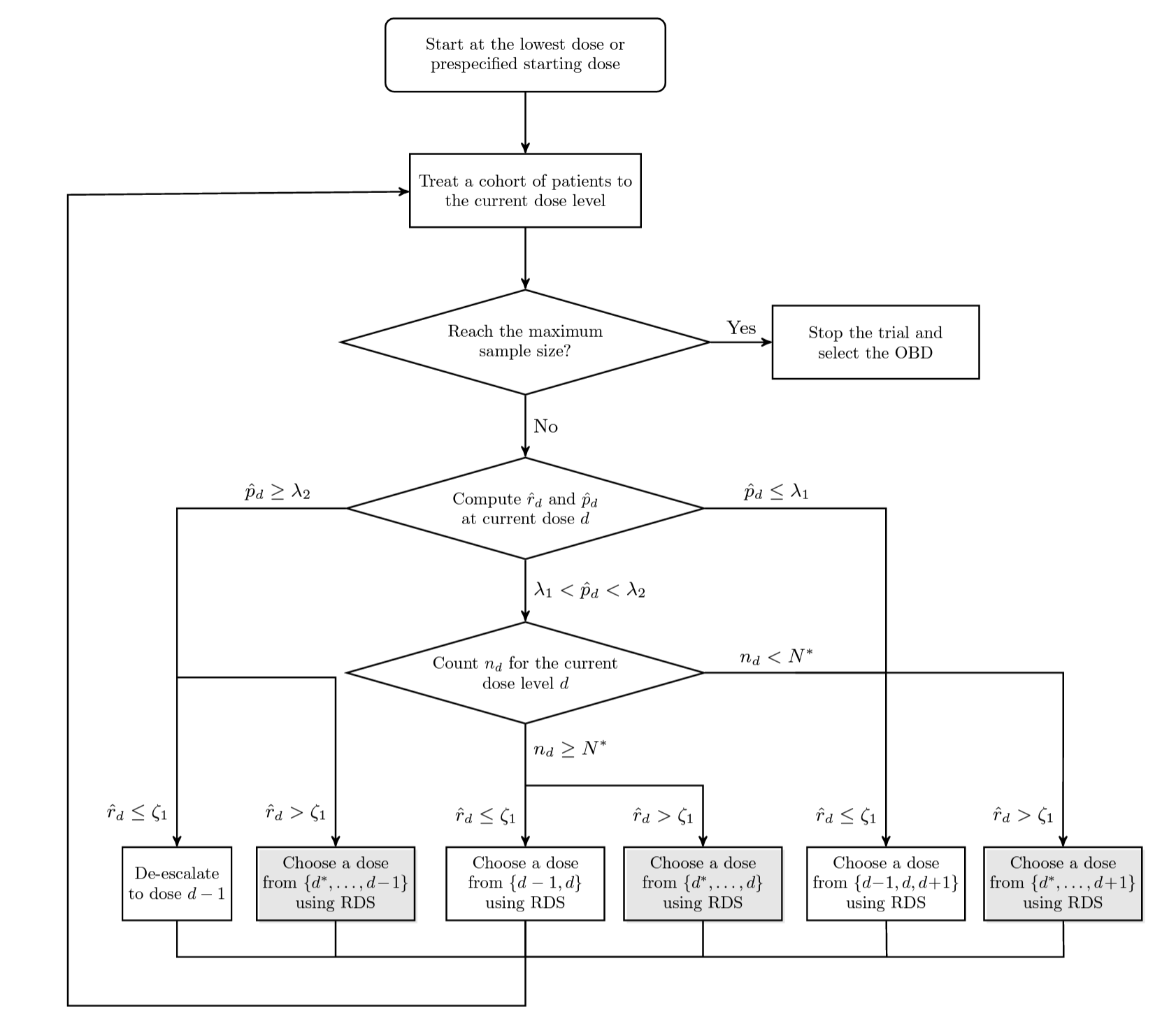}
    \caption{PKBOIN-12 flowchart}
    \label{fig:flowchart}
\end{figure}

Let the sample size cutoff $N^* = 6$, $c = 1$, and $d$ as the lowest or a pre-specified dose level. The dose-finding algorithm of the PKBOIN-12 design is shown in Figure~\ref{fig:flowchart} and listed below: 
\begin{itemize}
\item[1. ] Treat the cohort $c$ at the dose level $d$. 
\item[2. ] Derive the observed toxicity rate $\hat{p}_d$, the observed efficacy rate $\hat{q}_d$, and the observed sample mean of the PK value $\hat{r}_d$ for the current dose $d$. 
\item[3. ] If $\hat{r}_d \leq \zeta_1$, compare $\hat{p}_d$ with $\lambda_1$ and $\lambda_2$ to determine the following:
\begin{itemize}
    \item[(a)] for $\hat{p}_d \geq \lambda_2$, set $d'$ as the next lower dose level $d - 1$;
    \item[(b)] for $\hat{p}_d < \lambda_2$, $n_d \geq 9$, and the next higher dose level $d+1$ has not been used, set $d'$ to the next higher dose level $d+1$; 
    \item[(c)] for $\lambda_1 < \hat{p}_d < \lambda_2$, compare $n_d$ with $N^*$. When $n_d \geq N^*$, choose one dose $d'$ from the admissible dose set $A = \{d -1, d\}$ with larger RDS; otherwise choose one dose $d'$ from the admissible dose set $A = \{d -1, d, d+1 \}$ with the largest RDS;
    \item[(d)] for $\hat{p}_d \leq \lambda_1$, choose one dose from the admissible dose set $A = \{d-1 , d, d+1 \}$ with the largest RDS; 
\end{itemize}
\item[4. ] If $\hat{r}_d > \zeta_1$, compare $\hat{p}_d$ with $\lambda_1$ and $\lambda_2$ to determine the following:
\begin{itemize}
    \item[(a)] for $\hat{p}_d \geq \lambda_2$, choose one dose $d'$ from the admissible dose set $A = \{d^*, \ldots, d -1 \}$ with the largest RDS;
    \item[(b)] for $\hat{p}_d < \lambda_2$, $n_d \geq 9$, and the next higher dose level $d+1$ has not been used, set $d'$ to the next higher dose level $d+1$; 
    \item[(c)] for $\lambda_1 < \hat{p}_d < \lambda_2$, compare $n_d$ with $N^*$: When $n_d \geq N^*$, choose one dose $d'$ from the admissible dose set $A = \{d^*, \ldots, d\}$ with the largest RDS; otherwise choose one dose $d'$ from the admissible dose set $A = \{d^*, \ldots, d+1 \}$ with the largest RDS;
    \item[(d)] for $\hat{p}_d \leq \lambda_1$, choose one dose from the admissible dose set $A = \{d^*, \ldots, d+1 \}$ with the largest RDS.
\end{itemize}
\item[5. ] Set $c = c + 1$ and update $d = d'$ based on the dosing decision from Step 3 or Step 4. 
\item[6. ] Repeat Steps 1 - 5 until the maximum sample size is reached. 
\end{itemize}
When $\hat{r}_d \leq \zeta_1$, the dosing options in Step 3 are the same as those in Step 2 of BOIN12. However, if $\hat{r}_d > \zeta_1$, indicating sufficient PK exposure at the current dose level, the admissible set $A$ is expanded to include doses $\{d_{PK, min},  \ldots, d-2\}$, which have acceptable PK outcomes with lower toxicity probabilities. PKBOIN-12's design offers a more flexible set of admissible doses, accommodating various scenarios that can occur during a trial. For instance, if the current dose has an acceptable observed PK outcome and a high observed toxicity rate, i.e. $\hat{r}_d > \zeta_1$ and $\hat{p}_d \leq \lambda_2$, PKBOIN-12 chooses the dose with the highest RDS score from the range $\{d^*, \ldots, d - 1\}$. In a similar situation, BOIN12 would simply move to $d - 1$, which means that BOIN12 might need extra cohorts to allocate patients to the OBD. For example, if the OBD is $d - 2$, BOIN12 would require at least two cohorts to start treating new patients with the OBD.

PKBOIN-12 applies the same safety and efficacy criteria as BOIN12 to avoid allocating patients to toxic or inefficacious doses. Moreover, we specify another elimination criterion based on PK exposure, whereby dose levels with inefficacious PK outcomes are eliminated from consideration in subsequent dosing decisions. During the trial, 
\begin{itemize}[left=0em]
    \item[] \textbf{(PK)} \parbox[t]{0.9\linewidth}{if $\Pr(r_d < r_P \mid  \hat{r}_d, n_d) > C_P$ and $n_d \geq 6$, eliminate the lowest uneliminated dose level among $\{1, \ldots, d-1\}$ for $2 \leq d < D$ and eliminate all the dose levels and terminate the trial for $d = D$, }
\end{itemize}
where $C_P = 0.95$ denotes the probability cutoff for continuous PK value. This criterion serves as an additional method for eliminating dose levels that are not effective during the trial, which allows more cohorts to be allocated to the efficacious dose levels. Furthermore, if the highest dose level exhibits a high probability of having a PK exposure below the target PK outcome, all the dose levels will be considered ineffective and the trial will be terminated early.

\subsubsection{OBD Selection}\label{OBDselect}
Upon trial completion, we can estimate the toxicity, efficacy, and PK values for each dose level using the observed data. To ensure the monotonically increasing trend in toxicity and PK outcomes, we perform isotonic regression models to the observed toxicity probabilities $\{\hat{p}_d\}^D_{d=1}$ and PK sample means $\{\hat{r}_d\}^D_{d=1}$, utilizing the pool-adjacent-violators algorithm (PAVA) \cite{bril1984algorithm} to obtain the isotonically transformed values $\{\tilde{p}_d\}^D_{d=1}$ and $\{\tilde{r}_d\}^D_{d=1}$. Similar to BOIN12, we use these isotonic estimates of toxicity, $\{\tilde{p}_d\}^D_{d=1}$, to identify the MTD. However, unlike BOIN12 which defaults to selecting the first dose level as the lowest in the final admissible set, we integrate isotonic estimates of PK outcomes, $\{\tilde{r}_d\}^D_{d=1}$, to determine the lowest dose level for the final admissible set. PKBOIN-12 selects the final OBD through the following three steps: 
\begin{itemize}
    \item[(1)] determine the MTD by applying an isotonic regression to the observed toxicity rates $\{\hat{p}_d\}_{d=1}^D$ to obtain isotonic estimates $\{\tilde{p}_d\}_{d=1}^D$ and choose the dose level $d_{MTD}^*$ as MTD such that $d_{MTD}^* = \arg\min_d |\tilde{p}_d - p_T|$;
    \item[(2)] determine the minimum efficacious PK exposure dose $d_{PK, min}^*$ by applying an isotonic regression to the observed PK sample means $\{\hat{r}_d\}_{d=1}^D$ to obtain isotonic estimates $\{\tilde{r}_d\}_{d=1}^D$ such that $d_{PK, min}^* = \arg\min_d |\tilde{r}_d - r_P|$;
    \item[(3)] select one dose from $\{d_{PK, min}^*, \ldots, d_{MTD}\}$ as OBD with the highest estimated utility. 
\end{itemize}

Differing from the OBD selection procedure of BOIN12, our approach restricts the final list of OBD doses to only those levels that demonstrate efficacious PK outcomes. This strategy prevents the selection of an inefficacious dose level as the OBD. PKBOIN-12 employs a more restricted set for the final selection of the OBD, enhancing the probability of accurately identifying the OBD by eliminating doses that are inefficacious. Even in cases where PKBOIN-12 does not precisely identify the OBD, its more refined set of options can still enhance the likelihood of choosing a dose that closely aligns with the OBD. This advantage is illustrated in scenarios 5 and 13 of our simulation studies, where PKBOIN-12's effectiveness is demonstrated through the exclusion of doses that lack efficacy.

\subsection{Extension to TITE-PKBOIN-12 Design }\label{sec:TITE_PKBOIN_12}
Both BOIN12 and PKBOIN-12 utilize complete data from previous cohorts to determine the dosage for the subsequent patient cohort. However, the time required to assess toxicity and efficacy outcomes can span several months, potentially delaying the dosage decision for the next cohort. Some patients' outcomes in the previous cohorts may still be pending when the patients in the next cohort can be enrolled. In contrast, many PK outcomes can often be available within a few days after initial treatment. Typically, in clinical trials, the periods for assessing PK outcomes are considerably shorter than those for evaluating toxicity and efficacy outcomes. To make full use of the PK outcomes to shorten the trial duration, we propose the TITE-PKBOIN-12 design as an extension of PKBOIN-12. This extension is similar to TITE-BOIN12, which was introduced as an extension of BOIN12 to deal with the challenges of late-onset outcomes and rapid patient accrual  \cite{zhou2022tite}.

Let $A_T$ and $A_E$ be the toxicity and efficacy assessment time window, respectively. In the TITE-PKBOIN-12 design, which differs from BOIN12 and PKBOIN-12, we utilize the individual-level toxicity and efficacy data because each subject has a unique enrollment time. Denote $Y_{j,T}$ as the binary toxicity outcome and $Y_{j, E}$ as the binary efficacy outcome for the $j$th patient. Let $t$ be the time of the next dose assignment, $\delta_{j, T}(t)$ and $\delta_{j, E}(t)$ be the binary variables showing whether $Y_{j, T}$ and $Y_{j, E}$ have been ascertained ($\delta_{j, T}(t), \delta_{j, E}(t) = 1$) or are still pending ($\delta_{j, T}(t), \delta_{j, E}(t) = 0$) at time $t$. For example, if $\delta_{j,T}(t) = 1$, it indicates that $Y_{j,t}$ has been observed by time $t$; otherwise, $Y_{j,T}$ is still pending at time $t$. With the approximated likelihood (AL) approach from TITE-BOIN12 \cite{zhou2022tite}, the number of quasi-events of dose level $d$, $x_d$, can be approximated by $x_d^*(t)$ at time $t$ as follows: 
$$
x_d^*(t) = 100^{-1}\sum_{i=1}^4 u_in_{d,i}^*(t), 
$$
where $n_{d,i}^*(t)$ is an estimate of $n_{d,i}$ at time $t$ considering four possible missing patterns: (1) both toxicity and efficacy outcomes are not fully observed; (2) only toxicity outcome is fully observed; (3) only efficacy outcome is fully observed; (4) both toxicity and efficacy outcomes are fully observed. For simplicity, we ignore time point $t$ in the following notations and formulas. Similar to TITE-BOIN12, TITE-PKBOIN-12 assumes that $Y_{j,T}$ and $Y_{j, E}$ are independent. For each outcome $O_i$, $n_{d,i}^*$ is given by: 
$$\begin{aligned}
n_{d,i}^* = \sum_{j \in I_{d}} \Big\{ & I(Y_{j,T} = Y_{i, T})I(Y_{j,E} = Y_{i, T})\delta_{j, T}\delta_{j, E} + I(Y_{j,T} = Y_{i, T})E(Y_{j, E} = Y_{i, E}\mid \delta_{j, E} = 0)\delta_{j, T}(1 - \delta_{j, E}) \\
+ & E(Y_{j, T} = Y_{i, T}\mid \delta_{j, T} = 0)I(Y_{j,E} = Y_{i, E})(1 - \delta_{j, T})\delta_{j, E} \\
+ & E(Y_{j, T} = Y_{i, T}\mid \delta_{j, T} = 0)E(Y_{j, E} = Y_{i, E}\mid \delta_{j, E} = 0)(1 - \delta_{j, T})(1 - \delta_{j, E}) \Big\}, \quad i = 1, 2, 3, 4
\end{aligned}$$
where $(Y_{i, T}, Y_{i, E})$ are the toxicity and efficacy outcomes of $O_i$. Although the independence assumption is strong, TITE-PKBOIN-12 still has supreme performance in the OBD selection even when this assumption does not hold,  as evidenced in the sensitivity analysis presented in Section~\ref{sec:sensitivity}. The time to toxicity, $X_{j,T}$, and the time to efficacy, $X_{j, E}$, are assumed to follow the uniform distributions over two intervals, $[0, A_T]$ and $[0, A_E]$, respectively. Based on Yuan et al \cite{yuan2018time}, the two conditional expectations in the formula for $n_{d,i}^*$ can be calculated as follows:
$$
\begin{aligned}
E(Y_{j, T} = 1\mid \delta_{j, T} = 0) = \Pr(Y_{j, T} = 1\mid X_{j, T} > t_j) = \frac{\hat{p}_d^*(1 - t_j/A_T)}{1 - \hat{p}_d^* t_j/A_T}, \\ 
E(Y_{j, E} = 1\mid \delta_{j, E} = 0) = \Pr(Y_{j, E} = 1\mid X_{j, E} > t_j) = \frac{\hat{q}_d^*(1 - t_j/A_E)}{1 - \hat{q}_d^* t_j/A_E}, 
\end{aligned}
$$
where $\hat{p}_d^*$ and $\hat{q}_d^*$ are estimates of $p_d$ and $q_d$ with pending outcomes. Note that $X_{j,T}$ and $X_{j,E}$ could be modeled using different distributions, such as the Weibull distribution\cite{yuan2018time}. In such cases, the results for the two conditional expectations will change. The calculation details of $\hat{p}_d^*$ and $\hat{q}_d^*$ are presented in Section A2 of the supplementary document. With AL approach, we can estimate $\{n_{d,i}\}_{i=1}^4$, $\hat{p}_d$, $\hat{q}_d$, and $x_d$ by $\{n_{d,i}^*\}_{i=1}^4$, $\hat{p}_d^*$, $\hat{q}_d^*$, and $x_d^*$ for each dose level $d$. The detailed dose-finding algorithm of TITE-PKBOIN-12 is presented in Section A3 of the supplementary document. During the trial, TITE-PKBOIN-12 follows the four criteria below: 
\newlength{\secondlength}
\settowidth{\secondlength}{\textbf{(Efficacy)}}
\begin{itemize}[left=0em]
    \item[] \makebox[\secondlength][l]{\textbf{(Safety)}}   \parbox[t]{0.9\linewidth}{if $\Pr(p_d > p_T\mid \hat{p}_d^*, n_d) > C_T$, eliminate dose level $d$ and all above doses from the dose list;}
    \item[] \makebox[\secondlength][l]{\textbf{(Efficacy)}} \parbox[t]{0.9\linewidth}{if $\Pr(q_d < q_E\mid \hat{q}_d^*, n_d) > C_E$, eliminate dose level $d$ from the dose list;}
    \item[] \makebox[\secondlength][l]{\textbf{(PK)}}       \parbox[t]{0.9\linewidth}{if $\Pr(r_d < r_P \mid  \hat{r}_d, n_d) > C_P$ and $n_d \geq 6$, eliminate the lowest uneliminated dose level among $\{1, \ldots, d-1\}$ for $2 \leq d < D$ and eliminate all the dose levels and terminate the trial for $d = D$;}
    \item[] \makebox[\secondlength][l]{\textbf{(Accrual)}}  \parbox[t]{0.9\linewidth}{if more than 50\% of the patients have pending toxicity or efficacy outcomes at the current dose, suspend the accrual to wait for more data to become available.}
\end{itemize}
We adopt the same pre-specified values of $\{C_T, C_E, C_P\}$ as those used for PKBOIN-12. In the criteria for safety and efficacy, we substitute $\hat{p}_d$ and $\hat{q}_d$ with $\hat{p}_d^*$ and $\hat{q}_d^*$. The PK criterion remains unchanged from what is used in PKBOIN-12, as the PK outcome is presumed to be fully observed by the time of the next dosing decision. These criteria avoid allocating patients to toxic or inefficient doses. The last accrual criterion is consistent with the one utilized in TITE-BOIN12 \cite{zhou2022tite}, which avoids making risky decisions at the interim analysis when adequate information is not available. After the maximum sample size is reached and all toxicity and efficacy assessments have ended, TITE-PKBOIN-12 selects the final OBD following the same three steps introduced in Section~\ref{OBDselect}. Similar to PKBOIN-12, TITE-PKBOIN-12 limits the final OBD dose list to only include dose levels with efficacious PK outcomes. 

\section{Simulation}\label{sec:simulation}

\subsection{Scenario Configuration}
We conducted simulation studies to compare PKBOIN-12 and TITE-PKBOIN-12 with the existing BOIN12 and TITE-BOIN12 designs. We examined $D = 6$ dose levels, designating dose level 1 as the initial starting point. Each cohort contained three patients, with a total maximum sample size set at 45. Patient accrual occurred at a rate of three individuals per month. Both toxicity and efficacy outcomes were binary outcomes in the simulation studies. The windows for assessing toxicity and efficacy were set at $A_T = 30$ days and $A_E = 60$ days, respectively. The time-to-event variables, $X_{j,T}$ and $X_{j,E}$, were assumed to follow two uniform distributions over the intervals $[0, A_T]$ and $[0, A_E]$. We used AUC as the PK outcome for both PKBOIN-12 and TITE-PKBOIN-12. We consistently applied multiple common hyperparameters to ensure fair and balanced comparisons across all designs. We set the maximum acceptable toxicity probability $p_T = 0.35$, the lowest minimum acceptable efficacy probability $q_E = 0.25$, and assign the utility scores of $u_2 = 40$ for outcome $O_2$ and $u_3 = 60$ for outcome $O_3$. All the designs also shared the same hyperparameters $\lambda_1 = 0.276$, $\lambda_2 = 0.419$, $u_b = 0.705$, and RDS tables. The probability cutoffs for toxicity, efficacy, and PK were set at $C_T = 0.95$, $C_E = 0.9$, and $C_P = 0.95$, respectively. The accrual suspension rule in Section~\ref{sec:TITE_PKBOIN_12} was applied to both TITE-BOIN12 and TITE-PKBOIN-12 designs. 

In most of the early phase dosing finding simulation studies, the true toxicity and efficacy probabilities are constant for all patients assigned at the same dose level, i.e. $p_{d, j} = p_d$ and $q_{d,j} = q_d$, for each $j \in I_d$, so that we have $Y_{j, T} \sim \text{Bernoulli}(p_d)$ and $Y_{j, E} \sim \text{Bernoulli}(q_d)$. However, drug exposure can vary among patients at the same dose level. The AUC employed in our simulation studies serves as a measure of the overall drug exposure in the bloodstream over time. A higher AUC value indicates more drug accumulation in the body. The individual-level AUC values can have a large variability even at the same dose level. We considered the fact that the individual subjects have varying PK outcomes, which may influence their individual toxicity and efficacy responses. The individual-level toxicity and efficacy probabilities, $p_{d, j}$ and $q_{d, j}$, are linked with the actual PK outcome $r_{d, j}$ \cite{takeda2023bayesian}. We assumed the coefficient of variation (CV) for the individual-level PK outcome at 25\%. To ensure positive values, $r_{d, j}$ was generated using a truncated normal distribution, i.e. $r_{d,j} \sim \text{truncated-N}(r_d, (0.25 r_d)^2)$. With the actual individual-level PK values, the true individual-level toxicity and efficacy probabilities were assumed by 
$$
p_{d,j} = \min\{p_d(1 + g_P\frac{r_{d,j} - r_d}{r_d}), 1\}, \quad \text{and} \quad q_{d,j} = \min\{q_d(1+g_P\frac{r_{d,j} - r_d}{r_d}), 1\}, 
$$
where $g_P$ is a ratio that measures the relationship of the PK outcome with the true toxicity and efficacy probabilities at the individual level. A higher $g_P$ value implies that the variability in the drug exposure plays a more substantial role in affecting toxicity and efficacy outcomes. We set $g_P = 1$ in our simulations. Sensitivity analyses exploring various values of CV and $g_P$ are presented in Section~\ref{sec:sensitivity}. The toxicity and efficacy outcomes for each patient are generated based on their respective individual-level probabilities. Additional simulations examining the correlation between $Y_{j, T}$ and $Y_{j, E}$ are also discussed in Section~\ref{sec:sensitivity}.

\begin{figure}[htbp]
    \centering
    \includegraphics[width = 6in, height = 5in]{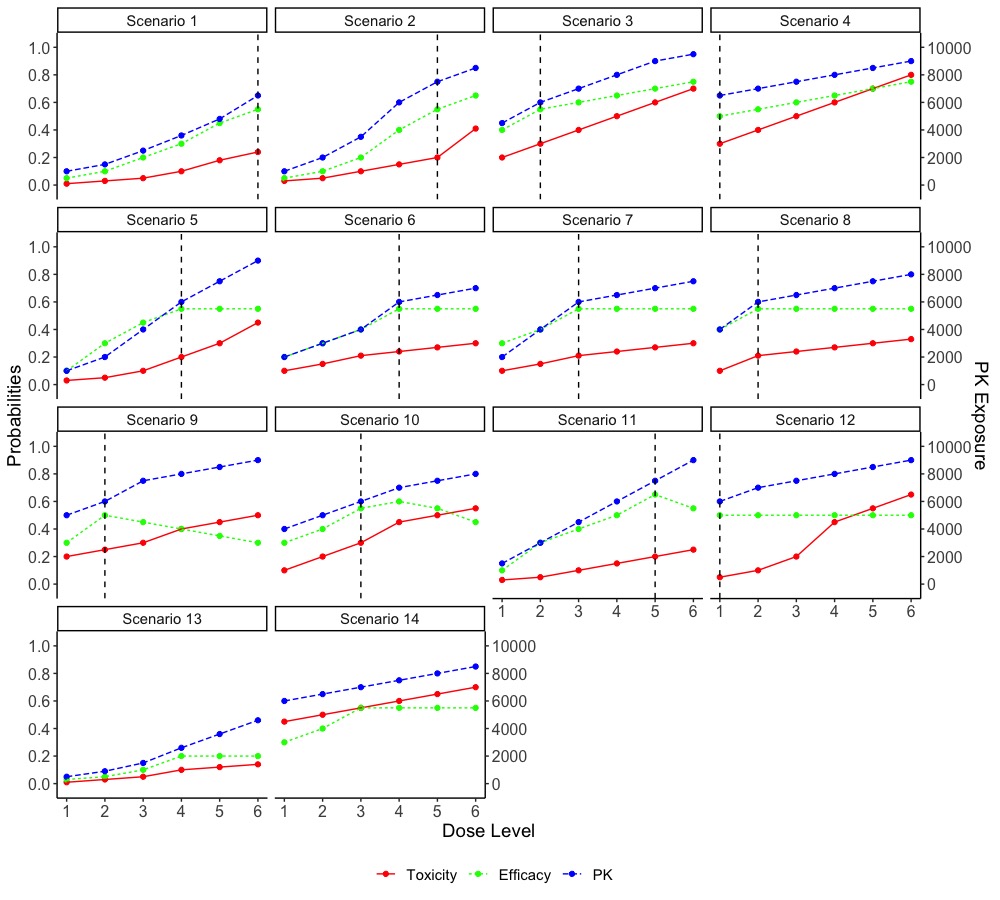}
    \caption{Curves of true toxicity probabilities, true efficacy probabilities, and true PK values under different scenarios}
    \label{fig:simu_truevalue}
\end{figure}

We considered 14 scenarios in our analysis. The toxicity and PK curves are presumed to increase monotonically. However, for the efficacy response, we explored four different dose-response relationships: 
\newlength{\lengthtwo}
\settowidth{\lengthtwo}{\textbf{Increasing: }}
\begin{itemize}[left=0em]
    \item[] \makebox[\lengthtwo][l]{\textbf{Increasing: }} \parbox[t]{0.9\linewidth}{response probabilities increase monotonically with dose levels;}
    \item[] \makebox[\lengthtwo][l]{\textbf{Plateau: }} \parbox[t]{1.9\linewidth}{response probabilities increase monotonically until a certain dose level, after which \\ they remain constant; }
    \item[] \makebox[\lengthtwo][l]{\textbf{Unimodal: }} \parbox[t]{0.9\linewidth}{response probabilities increase monotonically until a certain dose level, after which they decrease monotonically;}
    \item[] \makebox[\lengthtwo][l]{\textbf{Constant: }} \parbox[t]{0.9\linewidth}{response probabilities have the same value across all dose levels.}
\end{itemize}
The true PK values, along with the probabilities of toxicity and efficacy at each dose level across different scenarios, are shown in Figure~\ref{fig:simu_truevalue} and detailed in Table~\ref{tab:true_simuvalue}. The dose-response relationships and dose levels as the true OBD of all scenarios are presented in Table S2 of the supplementary document. In Figure~\ref{fig:simu_truevalue}, we use vertical dashed lines to highlight the true OBD in each scenario, which is also emphasized in bold in Table~\ref{tab:simuresult}. Scenarios 13 and 14 are two special cases when the true OBD does not exist if all dose levels are too toxic or inefficacious. The target AUC value was set at $r_P = 6000$. We conducted 2,000 independent replications for each scenario.

\subsection{Simulation Results}
The simulation results shown in Table~\ref{tab:simuresult} include both the proportion of selections as the OBD and the average number of patients assigned to each dose level. Additionally, the table outlines the proportion of early termination and averaged trial duration in months. Overall, compared to BOIN12 and TITE-BOIN12, PKBOIN-12 and TITE-PKBOIN-12 have higher probabilities of accurately identifying the OBD when the true OBD exists and terminating the trial when the true OBD does not exist. Our proposed designs also allocate more patients to the OBD while allocating much fewer patients to ineffective dose levels compared to BOIN12 and TITE-BOIN12. TITE-PKBOIN-12 has a substantially shorter trial duration than PKBOIN-12. The average trial durations for TITE-PKBOIN-12 and TITE-BOIN12 are nearly the same. However, in Scenario 13, TITE-PKBOIN-12 has a much shorter trial duration than TITE-BONI12. To prevent redundant comparisons, our comparison results primarily focus on highlighting the differences between PKBOIN-12 and BOIN12.

In the first four scenarios, we considered a monotonically increasing dose-response relationship with different dose levels designated as the true OBD. Scenario 1 set dose level 6 as the true OBD, which was also the only dose level with an effective PK value. PKBOIN-12 had an OBD selection probability of 53.8\%, which was 16.9\% higher than BOIN12. Dose levels 1, 2, and 3 were ineffective, with total selection probability values of around 13\% for BOIN12 and TITE-BOIN12, in contrast to a mere 0\% for PKBOIN-12 and TITE-PKBOIN-12. In Scenario 2, the true OBD was dose level 5. PKBOIN-12 had an OBD selection probability of 53.2\%, a 3.2\% improvement over BOIN12. Both BOIN12 and TITE-BOIN12 continued to have an approximately 10\% probability of selecting an ineffective dose level as the OBD. PKBOIN-12 has a 1.5\% lower probability of selecting the toxic dose level 6 compared to BOIN12. In Scenario 3, the true OBD was dose level 2. PKBOIN-12 still had a roughly 13.4\% greater chance of accurately selecting the OBD compared to BOIN12. In Scenario 4, dose level 1 served as the true OBD. All dose levels possessed effective PK values. All designs showed an approximate 70\% probability of correct OBD selection. PKBOIN-12 had a roughly 0.4\% higher probability of selecting the OBD relative to BOIN12.

In Scenarios 5 through 8, a plateau dose-response relationship was assumed once the target PK exposure level was reached. In Scenario 5, dose level 4 was identified as the OBD, while dose level 5 was the MTD. BOIN12 had an obviously lower probability of accurately selecting the OBD than PKBOIN-12, with 37.7\% versus 58.0\% respectively. Scenario 5 has the greatest discrepancy in OBD selection probabilities between BOIN12 and PKBOIN-12 when the true OBD existed. PKBOIN-12 exhibited a 19.7\% probability of selecting dose level 3 as the OBD. Conversely, BOIN12 had a 12.1\% probability of choosing either dose levels 1 or 2, and a 32.6\% probability of picking dose level 3 as the OBD. In Scenarios 6, 7, and 8, PKBOIN-12 consistently outperformed BOIN12 by having a greater probability of correctly selecting the OBD and lower probability values of choosing ineffective dose levels.

In Scenarios 9, 10, and 11, a unimodal dose-response relationship was considered. In Scenario 9, dose level 2 served as the OBD and exhibited the highest efficacy probability, all designs had over a 50\% probability of selecting the correct OBD. However, PKBOIN-12 had an OBD selection probability of 61.2\%, while BOIN12 had a slightly lower probability at 55.6\%. In Scenario 10, PKBOIN-12 displayed a 57.5\% probability of OBD selection, marking a substantial 19.1\% improvement compared to BOIN12. BOIN12 had a 23.8\% probability of selecting ineffective dose level 1, which had a lower toxicity probability but also a much lower efficacy probability compared to the OBD. However, PKBOIN-12 only had merely a 1\% probability of selecting that same dose. In Scenario 11, with dose level 5 as the OBD, PKBOIN-12 had a 45.1\% probability of OBD selection which was slightly higher than that of BOIN12. PKBOIN-12 displayed a 33.2\% probability of selecting the second-best dose, which was 12.0\% higher than BOIN12. 

Scenario 12 assumed a constant dose-response relationship making dose level 1 the true OBD. All designs exhibited roughly a 40\% probability of correctly identifying the OBD. PKBOIN-12 had a slightly higher probability of selecting the OBD and assigned more patients to the OBD than BOIN12. In Scenario 13, the OBD did not exist because all dose levels were ineffective. PKBOIN-12 and TITE-PKBOIN-12 showed about 85\% probability of terminating the trial early without selecting any dose levels. In contrast, both BOIN12 and TITE-BOIN12 had just merely about 0.3\% probability of early termination. Additionally, PKBOIN-12 had an average trial duration of 26.9 months, which was approximately 13 months shorter than that of BOIN12. PKBOIN-12 enrolled only about 10 cohorts of patients before ending the trial, whereas BOIN12 did not terminate early. Scenario 14 was another scenario without the true OBD because all dose levels were toxic. In this scenario, BOIN12 and PKBOIN-12 had a roughly similar probability of early termination, exceeding 50\%. However, BOIN12 had a 1.5\% higher probability of early termination compared to PKBOIN-12. The average trial duration in both designs was approximately 26.5 months.

Our simulation study revealed that PKBOIN-12 and TITE-PKBOIN-12 can effectively address the issue of a high probability of selecting an ineffective dose level as the OBD during the trial by incorporating PK information. This issue could result in no effective dose levels advancing to the dose-expansion phase or Phase II study, thereby causing the failure of the entire clinical development. In Scenarios 1, 2, 6, 11, and 13, where ineffective dose levels were present, both BOIN12 and TITE-BOIN12 displayed considerably higher probabilities of selecting these dose levels as the OBD. On the other hand, the probabilities of selecting toxic doses as the OBD were comparable across all four designs. Based on the simulation outcomes, our TITE-PKBOIN-12 design is most favorable, because of its high probability of accurate OBD selection, low risk of selecting ineffective doses, and shortest trial duration.

\subsection{Sensitivity Analysis}\label{sec:sensitivity}
We conducted a set of sensitivity analyses to assess the performance and robustness of both PKBOIN-12 and TITE-PKBOIN-12 designs. Firstly, we compared PKBOIN-12 with BOIN12 by varying the $g_P$ parameter through the values $\{0, 0.5, 1, 1.5, 2\}$. If $g_P = 0$, the PK outcome had no relationship with toxicity and efficacy outcomes at the individual level. Table~\ref{tab:sens_gP} presented the OBD selection probabilities of PKBOIN-12 and BOIN12, considering various $g_P$ values across all scenarios. PKBOIN-12 consistently outperformed BOIN12 for different $g_P$ values. Even in the cases where the PK outcome had no impact on toxicity and efficacy outcomes, i.e. $g_P = 0$, PKBOIN-12 still had superior performance. This is because PKBOIN-12 could exclude dose levels with ineffective PK outcomes from OBD selection. As $g_P$ increased, the OBD selection probability for PKBOIN-12 generally increased across the majority of scenarios. In the second sensitivity analysis, we adopted two alternative $CV$ values, 0.1 and 0.4, and presented the simulation results for BOIN12 and PKBOIN-12 in Tables S4 and S5 of the supplementary document. A high CV indicates substantial variability in individual-level PK outcomes. It was observed that PKBOIN-12 consistently outperformed BOIN12 across all scenarios with varying CV values. 

Additionally, we evaluated the robustness of PKBOIN-12 and TITE-PKBOIN-12 when the toxicity and efficacy outcomes were correlated. We conducted two separate simulation studies given the correlation coefficients between toxicity and efficacy outcomes as $\rho_{pq} = 0.2$ and $\rho_{pq} = 0.4$. The outcomes are presented in Table S6 and Table S7, respectively. Despite the design algorithms of PKBOIN-12 and TITE-PKBOIN-12 operating under the independence assumption between toxicity and efficacy outcomes, both designs demonstrated robustness across all examined scenarios. When there was a positive correlation between toxicity and efficacy, the OBD selection probabilities for our designs tended to increase slightly in the majority of the scenarios. We further examined the performance of two proposed designs under different utility score settings, specifically $(u_{11}, u_{00}) = (100, 0)$. This setting focused on identifying the dose level with the highest efficacy probability while keeping the toxicity probability not greater than $p_T$. As illustrated in Table S8 of the supplementary material, both PKBOIN-12 and TITE-PKBOIN-12 maintained similar probabilities for OBD selection across all scenarios. Finally, we assessed the performance of the two designs with longer assessment windows, denoted as $(A_T, A_E) = \{(45, 90), (90, 120)\}$.  From Table S9 and Table S10, both designs continued to exhibit robust probabilities for OBD selection and patient assignments to OBDs with different assessment windows. With longer assessment windows, TITE-PKBOIN-12 demonstrated a greater ability to reduce trial length. Compared with PKBOIN-12, TITE-PKBOIN-12 reduced the trial duration by approximately 10.2 months for $(A_T, A_E) = (30, 60)$, 14.9 months for $(A_T, A_E) = (45, 90)$, and 19.0 months for $(A_T, A_E) = (90, 120)$.

\section{Software}\label{sec:software}
The software for implementing BOIN12 and TITE-BOIN12 are available at \href{https://trialdesign.org/}{www.trialdesign.org}. Our proposed PKBOIN-12 and TITE-PKBOIN-12 designs are shared at \href{https://github.com/EugeneHao/PKBOIN-12}{https://github.com/EugeneHao/PKBOIN-12}.

\section{Conclusion}\label{sec:conclusion}
Early phase dose-finding clinical trials often collect PK outcomes as they serve as valuable indicators for assessing the initial safety and efficacy of the study drug based on exposure-safety and exposure-efficacy analyses. We developed the innovative PKBOIN-12 design, which integrates PK data with toxicity and efficacy outcomes to identify the optimal dosage. Additionally, to address the issue of late-onset toxicity and efficacy outcomes, we introduced the TITE-PKBOIN-12 design, which is advantageous when the PK outcomes are available much quicker than toxicity and efficacy outcomes. Both PKBOIN-12 and TITE-PKBOIN-12 are extensions of BOIN-12 and can utilize utility scores and the RDS table for OBD selection, enhancing the risk-benefit trade-off. Similar to TITE-BOIN12, TITE-PKBOIN-12 employs an approximated likelihood method for estimating the unobserved count of quasi-events. Thus, both PKBOIN-12 and TITE-PKBOIN-12 are straightforward to implement in clinical trials, requiring only the additional PK outcome for dosing decisions.

In our simulation studies, we explored four different dose-response relationship patterns. Our simulations also considered the variability in individual outcomes due to the varying levels of drug exposure. The simulation results demonstrated that PKBOIN-12 improved the probability of selecting the OBD in comparison to BOIN12, and it also decreased the chances of selecting ineffective dose levels. Additionally, PKBOIN-12 was more effective in assigning patients to the optimal dose than BOIN12. Moreover, our sensitivity analysis revealed that PKBOIN-12 and TITE-PKBOIN-12 had robust performance. PKBOIN-12 consistently outperformed BOIN12 under varying conditions of individual PK outcome and differing associations between PK outcome and toxicity and efficacy outcomes. 

Our proposed designs, PKBOIN-12 and TITE-PKBOIN-12, have certain limitations. Firstly, they only incorporate a single PK outcome. In cases where multiple PK outcomes are available in a clinical trial, it is crucial to collaborate with physicians and pharmacometricians to determine which PK outcome should be utilized, as this choice can influence the OBD selection result. Secondly, we presuppose a monotonic relationship between the dose level and the chosen PK outcome in our designs. While this assumption holds for many PK variables, it may not apply universally. Thus, a thorough analysis of the dose-exposure relationship is necessary before selecting the PK outcome for the designs. Thirdly, we assume immediate availability of PK data for dosing decisions for subsequent cohorts. However, in practical clinical trial settings, biostatisticians often face delays in accessing PK data due to the extensive data collection and processing protocols, which aim to maintain data integrity and accuracy. Therefore, the reduction in trial duration offered by TITE-PKBOIN-12 might not be as substantial in actual clinical trials as our simulation results.

There are several ways to further extend our proposed designs. First, it is feasible to substitute the continuous PK outcome with a binary one. Moreover, our designs can be expanded to include multiple PK outcomes. One method is to create a composite PK outcome by combining several PK outcomes using the Mahalanobis distance. Alternatively, a weighted average of the PK outcomes could be employed. This adaptation is particularly beneficial in scenarios where a single PK outcome does not adequately represent the exposure-safety or exposure-efficacy relationship. Another extension is to utilize the categorical toxicity and efficacy outcomes. While our discussion focused on binary outcomes, similar to BOIN12 and TITE-BOIN12, our proposed designs can be adapted to accommodate categorical toxicity and efficacy outcomes by developing a new utility table as the combination of the two categorical variables. The utility values are determined in consultation with physicians. The last extension is to incorporate a nonlinear influence function of the PK outcome on the toxicity and efficacy outcomes. Distinct influence functions could be applied to toxicity and efficacy, allowing for a more accurate reflection of how PK exposure relates to both outcomes.

\section*{Declaration of Interest Statement}

The authors declare that they have no known competing financial interests or personal relationships that could have appeared to influence the work reported in this paper.

\section*{Data Availability}
This paper does not use any real data. The simulation code will be available on request.

\bibliographystyle{plain}
\bibliography{cite}

\clearpage

\begin{table}[htbp]
\caption{True toxicity probabilities, efficacy probabilities, and PK values for each dose level}\label{tab:true_simuvalue}
\centering
\begin{tabular}{|l|cccccc|cccccc|}
\hline
Category & \multicolumn{12}{c|}{Dose Level}\\
\hline
  & 1 & 2 & 3 & 4 & 5 & 6 & 1 & 2 & 3 & 4 & 5 & 6 \\
\hline
&  \multicolumn{6}{c|}{Scenario 1} & \multicolumn{6}{c|}{Scenario 2} \\
Toxicity & 0.01 & 0.03 & 0.05 & 0.10 & 0.18 & \textbf{0.24} & 0.03 & 0.05 & 0.10 & 0.15 & \textbf{0.20} & 0.41 \\
Efficacy & 0.05 & 0.10 & 0.20 & 0.30 & 0.45 & \textbf{0.55} & 0.05 & 0.10 & 0.20 & 0.40 & \textbf{0.55} & 0.65 \\
PK       & 1000 & 1500 & 2500 & 3600 & 4800 & \textbf{6500} & 1000 & 2000 & 3500 & 6000 & \textbf{7500} & 8500 \\
\hline
& \multicolumn{6}{c|}{Scenario 3} & \multicolumn{6}{c|}{Scenario 4} \\ 
Toxicity & 0.20 & \textbf{0.30} & 0.40 & 0.50 & 0.60 & 0.70 & \textbf{0.30} & 0.40 & 0.50 & 0.60 & 0.70 & 0.80 \\
Efficacy & 0.40 & \textbf{0.55} & 0.60 & 0.65 & 0.70 & 0.75 & \textbf{0.50} & 0.55 & 0.60 & 0.65 & 0.70 & 0.75 \\
PK       & 4500 & \textbf{6000} & 7000 & 8000 & 9000 & 9500 & \textbf{6500} & 7000 & 7500 & 8000 & 8500 & 9000 \\
\hline
& \multicolumn{6}{c|}{Scenario 5} & \multicolumn{6}{c|}{Scenario 6} \\
Toxicity & 0.03 & 0.05 & 0.10 & \textbf{0.20} & 0.30 & 0.45 & 0.10 & 0.15 & 0.21 & \textbf{0.24} & 0.27 & 0.30 \\
Efficacy & 0.10 & 0.30 & 0.45 & \textbf{0.55} & 0.55 & 0.55 & 0.20 & 0.30 & 0.40 & \textbf{0.55} & 0.55 & 0.55 \\
PK       & 1000 & 2000 & 4000 & \textbf{6000} & 7500 & 9000 & 2000 & 3000 & 4000 & \textbf{6000} & 6500 & 7000 \\
\hline
& \multicolumn{6}{c|}{Scenario 7} & \multicolumn{6}{c|}{Scenario 8} \\
Toxicity & 0.10 & 0.15 & \textbf{0.21} & 0.24 & 0.27 & 0.30 & 0.10 & \textbf{0.21} & 0.24 & 0.27 & 0.30 & 0.33 \\
Efficacy & 0.30 & 0.40 & \textbf{0.55} & 0.55 & 0.55 & 0.55 & 0.40 & \textbf{0.55} & 0.55 & 0.55 & 0.55 & 0.55 \\
PK       & 2000 & 4000 & \textbf{6000} & 6500 & 7000 & 7500 & 4000 & \textbf{6000} & 6500 & 7000 & 7500 & 8000 \\
\hline
& \multicolumn{6}{c|}{Scenario 9} & \multicolumn{6}{c|}{Scenario 10} \\
Toxicity & 0.20 & \textbf{0.25} & 0.30 & 0.40 & 0.45 & 0.50 & 0.10 & 0.20 & \textbf{0.30} & 0.45 & 0.50 & 0.55  \\
Efficacy & 0.30 & \textbf{0.50} & 0.45 & 0.40 & 0.35 & 0.30 & 0.30 & 0.40 & \textbf{0.55} & 0.60 & 0.55 & 0.45 \\
PK       & 5000 & \textbf{6500} & 7500 & 8000 & 8500 & 9000 & 4000 & 5000 & \textbf{6000} & 7000 & 7500 & 8000 \\
\hline
& \multicolumn{6}{c|}{Scenario 11} & \multicolumn{6}{c|}{Scenario 12} \\
Toxicity & 0.03 & 0.05 & 0.10 & 0.15 & \textbf{0.20} & 0.25 & \textbf{0.05} & 0.10 & 0.20 & 0.45 & 0.55 & 0.65 \\
Efficacy & 0.10 & 0.30 & 0.40 & 0.50 & \textbf{0.65} & 0.55 & \textbf{0.50} & 0.50 & 0.50 & 0.50 & 0.50 & 0.50 \\
PK       & 1500 & 3000 & 4500 & 6000 & \textbf{7500} & 9000 & \textbf{6000} & 7000 & 7500 & 8000 & 8500 & 9000 \\
\hline
& \multicolumn{6}{c|}{Scenario 13} & \multicolumn{6}{c|}{Scenario 14} \\
Toxicity & 0.01 & 0.03 & 0.05 & 0.10 & 0.12 & 0.14 & 0.45 & 0.50 & 0.55 & 0.60 & 0.65 & 0.70 \\
Efficacy & 0.03 & 0.05 & 0.10 & 0.20 & 0.20 & 0.20 & 0.30 & 0.40 & 0.55 & 0.55 & 0.55 & 0.55 \\
PK       & 500  & 900  & 1500 & 2600 & 3600 & 4600 & 6000 & 6500 & 7000 & 7500 & 8000 & 8500 \\
\hline
\end{tabular}
\end{table}

\clearpage

\setlength{\LTleft}{-1cm} 
\begin{longtable}{|l|ccccccc|cccccc|c|}
\caption{Simulation results for all designs with $p_T = 0.35$, $q_E = 0.25$, including selection probability for each dose level, number of patients assigned to each dose level, and average trial duration}\label{tab:simuresult}\\
\hline
\multirow{2}{*}{} & \multicolumn{7}{c|}{Selection Probability (\%)} & \multicolumn{6}{c|}{Number of Assigned Patients} & Duration\\
\cline{2-15}
  & 1 & 2 & 3 & 4 & 5 & 6 & ET & 1 & 2 & 3 & 4 & 5 & 6 &  Months\\
\hline
Design & \multicolumn{14}{c|}{Scenario 1} \\
\hline
BOIN12 & 1.7 & 2.5 & 8.9 & 19.1 & 30.8 & \textbf{36.9} & 0.1 & 3.8 & 4.4 & 5.8 & 8.2 & 11.5 & \textbf{11.4} & 38.4 \\ 
PKBOIN-12 & 0.0 & 0.0 & 0.1 & 3.3 & 40.9 & \textbf{53.8} & 1.9 & 3.0 & 3.3 & 5.1 & 8.2 & 12.6 & \textbf{12.6} & 38.1 \\ 
TITE-BOIN12 & 1.3 & 2.5 & 8.6 & 18.3 & 32.0 & \textbf{37.2} & 0.1 & 3.7 & 4.4 & 6.0 & 8.4 & 11.5 & \textbf{10.9} & 25.2 \\ 
TITE-PKBOIN12 & 0.0 & 0.0 & 0.0 & 3.2 & 41.9 & \textbf{52.8} & 2.0 & 3.2 & 3.6 & 5.4 & 8.5 & 12.7 & \textbf{11.4} & 25.0 \\ 
\hline
Design & \multicolumn{14}{c|}{Scenario 2} \\
\hline
BOIN12 & 1.6 & 2.5 & 6.0 & 22.7 & \textbf{50.0} & 17.2 & 0.1 & 3.9 & 4.5 & 5.5 & 9.5 & \textbf{14.5} & 7.0 & 37.9 \\ 
PKBOIN-12 & 0.0 & 0.0 & 2.4 & 28.5 & \textbf{53.2} & 15.7 & 0.3 & 3.1 & 4.1 & 6.1 & 10.9 & \textbf{14.2} & 6.6 & 37.9 \\ 
TITE-BOIN12 & 1.2 & 2.4 & 6.3 & 23.8 & \textbf{47.9} & 18.2 & 0.1 & 4.0 & 4.5 & 5.7 & 10.0 & \textbf{13.8} & 6.9 & 24.7 \\ 
TITE-PKBOIN12 & 0.0 & 0.0 & 2.5 & 28.1 & \textbf{52.0} & 17.1 & 0.3 & 3.3 & 4.2 & 6.1 & 11.0 & \textbf{13.9} & 6.5 & 24.7 \\ 
\hline
Design & \multicolumn{14}{c|}{Scenario 3} \\
\hline
BOIN12 & 35.9 & \textbf{45.9} & 14.3 & 2.3 & 0.0 & 0.0 & 1.5 & 17.3 & \textbf{17.5} & 7.5 & 1.9 & 0.3 & 0.0 & 36.7 \\ 
PKBOIN-12 & 18.7 & \textbf{59.2} & 17.0 & 2.4 & 0.2 & 0.0 & 2.4 & 16.2 & \textbf{18.6} & 7.4 & 1.8 & 0.3 & 0.0 & 36.5 \\ 
TITE-BOIN12 & 35.9 & \textbf{44.3} & 15.3 & 2.5 & 0.2 & 0.0 & 1.7 & 17.0 & \textbf{16.9} & 7.8 & 2.3 & 0.4 & 0.0 & 20.8 \\ 
TITE-PKBOIN12 & 19.4 & \textbf{59.2} & 16.1 & 2.8 & 0.2 & 0.0 & 2.2 & 16.1 & \textbf{18.3} & 7.6 & 2.0 & 0.4 & 0.0 & 20.7 \\
\hline
Design & \multicolumn{14}{c|}{Scenario 4} \\ 
\hline
BOIN12 & \textbf{69.8} & 19.2 & 2.5 & 0.1 & 0.0 & 0.0 & 8.3 & \textbf{27.8} & 10.8 & 3.1 & 0.5 & 0.0 & 0.0 & 34.8 \\ 
PKBOIN-12 & \textbf{70.2} & 18.2 & 2.6 & 0.2 & 0.0 & 0.0 & 8.7 & \textbf{28.3} & 10.5 & 2.9 & 0.5 & 0.1 & 0.0 & 34.9 \\ 
TITE-BOIN12 & \textbf{69.4} & 20.6 & 2.8 & 0.5 & 0.0 & 0.0 & 6.6 & \textbf{27.0} & 11.5 & 3.5 & 0.7 & 0.1 & 0.0 & 19.4 \\ 
TITE-PKBOIN12 & \textbf{70.4} & 19.7 & 2.8 & 0.2 & 0.0 & 0.0 & 6.8 & \textbf{28.0} & 10.9 & 3.2 & 0.7 & 0.1 & 0.0 & 19.3 \\ 
\hline
Design & \multicolumn{14}{c|}{Scenario 5} \\
\hline
BOIN12 & 1.0 & 11.1 & 32.6 & \textbf{37.7} & 15.3 & 2.1 & 0.0 & 3.7 & 7.3 & 12.6 & \textbf{12.6} & 6.6 & 2.1 & 37.7 \\ 
PKBOIN-12 & 0.0 & 0.0 & 19.7 & \textbf{58.0} & 19.2 & 2.9 & 0.1 & 3.2 & 6.3 & 13.3 & \textbf{13.8} & 6.5 & 1.9 & 37.6 \\ 
TITE-BOIN12 & 0.6 & 10.9 & 31.7 & \textbf{37.9} & 16.2 & 2.7 & 0.0 & 3.9 & 7.8 & 12.7 & \textbf{12.3} & 6.4 & 2.0 & 23.4 \\ 
TITE-PKBOIN12 & 0.0 & 0.0 & 18.4 & \textbf{58.2} & 20.3 & 2.8 & 0.1 & 3.5 & 7.2 & 13.1 & \textbf{12.9} & 6.4 & 1.9 & 23.4 \\ 
\hline
Design & \multicolumn{14}{c|}{Scenario 6} \\
\hline
BOIN12 & 7.3 & 14.4 & 18.7 & \textbf{34.0} & 18.6 & 6.7 & 0.3 & 7.1 & 8.9 & 9.2 & \textbf{10.7} & 6.2 & 2.7 & 37.9 \\ 
PKBOIN-12 & 0.1 & 1.7 & 17.2 & \textbf{47.1} & 21.6 & 9.7 & 2.5 & 4.6 & 7.9 & 10.7 & \textbf{11.9} & 6.3 & 3.0 & 37.4 \\ 
TITE-BOIN12 & 6.9 & 14.1 & 17.2 & \textbf{35.7} & 17.7 & 8.1 & 0.3 & 7.3 & 9.0 & 9.2 & \textbf{10.6} & 6.0 & 2.8 & 23.4 \\ 
TITE-PKBOIN12 & 0.2 & 1.5 & 16.6 & \textbf{46.1} & 23.4 & 9.9 & 2.3 & 5.1 & 8.0 & 10.8 & \textbf{11.5} & 6.3 & 2.8 & 23.4 \\ 
\hline
Design & \multicolumn{14}{c|}{Scenario 7} \\
\hline
BOIN12 & 9.1 & 17.0 & \textbf{34.8} & 23.2 & 11.3 & 4.5 & 0.1 & 7.9 & 10.0 & \textbf{12.9} & 8.2 & 4.1 & 1.8 & 37.5 \\ 
PKBOIN-12 & 0.2 & 10.2 & \textbf{45.1} & 26.0 & 12.4 & 4.9 & 1.0 & 6.0 & 10.8 & \textbf{14.3} & 8.1 & 3.9 & 1.6 & 37.3 \\ 
TITE-BOIN12 & 9.0 & 18.2 & \textbf{35.9} & 20.7 & 10.8 & 5.2 & 0.2 & 8.4 & 10.7 & \textbf{13.0} & 7.6 & 3.6 & 1.7 & 22.6 \\ 
TITE-PKBOIN12 & 0.4 & 10.6 & \textbf{46.3} & 25.1 & 11.7 & 5.3 & 0.7 & 6.7 & 11.1 & \textbf{14.1} & 7.9 & 3.5 & 1.5 & 22.5 \\ 
\hline
Design & \multicolumn{14}{c|}{Scenario 8} \\
\hline
BOIN12 & 20.2 & \textbf{37.9} & 21.4 & 12.2 & 5.6 & 2.6 & 0.2 & 12.0 & \textbf{15.9} & 8.9 & 4.7 & 2.2 & 1.1 & 37.1 \\ 
PKBOIN-12 & 8.7 & \textbf{47.0} & 25.3 & 12.3 & 4.8 & 1.8 & 0.1 & 11.8 & \textbf{17.0} & 9.3 & 4.4 & 1.7 & 0.7 & 37.1 \\ 
TITE-BOIN12 & 19.2 & \textbf{37.6} & 22.8 & 12.7 & 5.2 & 2.3 & 0.2 & 12.5 & \textbf{15.8} & 9.0 & 4.7 & 2.1 & 0.9 & 21.6 \\ 
TITE-PKBOIN12 & 8.8 & \textbf{48.3} & 24.3 & 12.2 & 4.6 & 1.7 & 0.2 & 12.2 & \textbf{17.1} & 9.2 & 4.2 & 1.6 & 0.6 & 21.5 \\ 
\hline
Design & \multicolumn{14}{c|}{Scenario 9} \\
\hline
BOIN12 & 18.0 & \textbf{55.6} & 19.7 & 3.8 & 0.9 & 0.1 & 2.0 & 12.3 & \textbf{18.0} & 8.8 & 3.4 & 1.4 & 0.5 & 37.4 \\ 
PKBOIN-12 & 9.8 & \textbf{61.2} & 20.8 & 4.4 & 0.5 & 0.2 & 3.0 & 12.6 & \textbf{18.8} & 8.4 & 3.0 & 1.0 & 0.4 & 37.3 \\ 
TITE-BOIN12 & 17.3 & \textbf{54.5} & 20.8 & 4.3 & 0.9 & 0.2 & 1.9 & 12.1 & \textbf{17.9} & 9.0 & 3.5 & 1.4 & 0.5 & 22.1 \\ 
TITE-PKBOIN12 & 9.8 & \textbf{62.2} & 20.4 & 4.2 & 0.8 & 0.1 & 2.5 & 12.7 & \textbf{18.7} & 8.5 & 3.1 & 1.0 & 0.3 & 21.9 \\ 
\hline
Design & \multicolumn{14}{c|}{Scenario 10} \\
\hline
BOIN12 & 23.8 & 28.2 & \textbf{38.4} & 8.8 & 0.5 & 0.0 & 0.3 & 12.2 & 13.3 & \textbf{13.3} & 5.0 & 1.0 & 0.2 & 37.7 \\ 
PKBOIN-12 & 0.9 & 25.9 & \textbf{57.5} & 12.7 & 1.2 & 0.1 & 1.8 & 8.3 & 15.1 & \textbf{14.9} & 5.2 & 1.0 & 0.2 & 37.3 \\ 
TITE-BOIN12 & 22.2 & 28.6 & \textbf{38.4} & 9.4 & 1.0 & 0.0 & 0.3 & 11.8 & 13.3 & \textbf{13.2} & 5.3 & 1.1 & 0.2 & 22.1 \\ 
TITE-PKBOIN12 & 1.2 & 25.9 & \textbf{57.0} & 12.7 & 1.6 & 0.1 & 1.5 & 8.7 & 14.9 & \textbf{14.5} & 5.3 & 1.2 & 0.2 & 22.0 \\ 
\hline
Design & \multicolumn{14}{c|}{Scenario 11} \\
\hline
BOIN12 & 0.4 & 10.1 & 15.0 & 21.2 & \textbf{43.1} & 10.2 & 0.0 & 3.7 & 7.2 & 8.4 & 9.9 & \textbf{11.6} & 4.2 & 37.5 \\ 
PKBOIN-12 & 0.0 & 0.0 & 10.4 & 33.2 & \textbf{45.1} & 10.9 & 0.2 & 3.2 & 6.1 & 9.7 & 10.6 & \textbf{11.3} & 4.1 & 37.5 \\ 
TITE-BOIN12 & 0.4 & 9.8 & 15.6 & 22.6 & \textbf{42.9} & 8.8 & 0.0 & 3.9 & 7.7 & 9.2 & 10.1 & \textbf{10.6} & 3.6 & 23.7 \\ 
TITE-PKBOIN12 & 0.0 & 0.0 & 12.3 & 33.4 & \textbf{43.6} & 10.5 & 0.1 & 3.5 & 7.0 & 10.3 & 10.4 & \textbf{10.2} & 3.6 & 23.7 \\ 
\hline
Design & \multicolumn{14}{c|}{Scenario 12} \\
\hline
BOIN12 & \textbf{41.6} & 37.2 & 19.0 & 1.8 & 0.3 & 0.0 & 0.0 & \textbf{17.2} & 14.8 & 9.0 & 3.0 & 0.8 & 0.1 & 37.3 \\ 
PKBOIN-12 & \textbf{44.5} & 36.5 & 17.0 & 1.8 & 0.1 & 0.0 & 0.0 & \textbf{18.6} & 15.3 & 8.0 & 2.5 & 0.5 & 0.1 & 37.3 \\ 
TITE-BOIN12 & \textbf{43.6} & 34.3 & 19.7 & 2.1 & 0.2 & 0.0 & 0.0 & \textbf{18.1} & 14.8 & 8.6 & 2.7 & 0.7 & 0.2 & 21.1 \\ 
TITE-PKBOIN12 & \textbf{45.1} & 35.5 & 17.2 & 2.1 & 0.1 & 0.0 & 0.0 & \textbf{20.1} & 14.7 & 7.4 & 2.2 & 0.5 & 0.1 & 21.0 \\
\hline
Design & \multicolumn{14}{c|}{Scenario 13} \\
\hline
BOIN12 & 6.9 & 8.3 & 13.7 & 31.5 & 23.8 & 15.5 & \textbf{0.3} & 5.2 & 6.1 & 7.1 & 9.6 & 9.5 & 7.5 & 39.8 \\ 
PKBOIN-12 & 0.0 & 0.0 & 0.0 & 0.5 & 4.8 & 10.0 & \textbf{84.7} & 3.0 & 3.3 & 4.3 & 6.7 & 7.2 & 5.8 & 26.9 \\ 
TITE-BOIN12 & 5.8 & 6.5 & 13.9 & 31.8 & 25.2 & 16.8 & \textbf{0.2} & 5.0 & 5.8 & 6.9 & 9.7 & 9.8 & 7.9 & 27.5 \\ 
TITE-PKBOIN12 & 0.0 & 0.0 & 0.0 & 0.7 & 4.0 & 10.2 & \textbf{85.0} & 3.1 & 3.4 & 4.4 & 6.9 & 7.3 & 5.8 & 20.8 \\ 
\hline
Design & \multicolumn{14}{c|}{Scenario 14} \\
\hline
BOIN12 & 39.0 & 3.8 & 0.8 & 0.2 & 0.0 & 0.0 & \textbf{56.3} & 22.9 & 5.5 & 1.8 & 0.4 & 0.0 & 0.0 & 26.3 \\ 
PKBOIN-12 & 40.5 & 3.7 & 0.9 & 0.1 & 0.0 & 0.0 & \textbf{54.8} & 23.1 & 5.5 & 1.8 & 0.3 & 0.0 & 0.0 & 26.5 \\ 
TITE-BOIN12 & 45.8 & 3.3 & 0.6 & 0.1 & 0.0 & 0.0 & \textbf{50.2} & 24.3 & 6.0 & 1.9 & 0.4 & 0.1 & 0.0 & 15.9 \\ 
TITE-PKBOIN12 & 44.5 & 4.6 & 0.9 & 0.1 & 0.0 & 0.0 & \textbf{49.9} & 24.2 & 6.0 & 2.0 & 0.5 & 0.1 & 0.0 & 15.9 \\
\hline
\end{longtable}

\begin{table}[ht]
\caption{OBD selection probabilities for BOIN12 (B) and PKBOIN-12 (P) across all scenarios given different values of $g_P$}\label{tab:sens_gP}
\centering
\begin{tabular}{|c|cc|cc|cc|cc|cc|}
\hline
Scenario & \multicolumn{2}{c|}{$g_P = 0$} & \multicolumn{2}{c|}{$g_P = 0.5$} & \multicolumn{2}{c|}{$g_P = 1$} & \multicolumn{2}{c|}{$g_P = 1.5$} & \multicolumn{2}{c|}{$g_P = 2$} \\ 
\hline
& B & P & B & P & B & P & B & P & B & P \\
\hline
1 & 37.1 & 51.6 & 37.4 & 51.8 & 36.9 & 53.8 & 38.2 & 56.0 & 37.4 & 57.0 \\ 
2 & 49.6 & 51.4 & 45.6 & 52.3 & 50.0 & 53.2 & 51.6 & 51.5 & 50.6 & 52.7 \\ 
3 & 46.1 & 58.2 & 44.8 & 58.3 & 45.9 & 59.2 & 46.5 & 57.2 & 46.5 & 58.3 \\ 
4 & 69.7 & 70.1 & 69.8 & 69.3 & 69.8 & 70.2 & 69.2 & 70.8 & 70.8 & 70.3 \\ 
5 & 40.5 & 56.9 & 38.0 & 56.6 & 37.7 & 58.0 & 37.2 & 58.0 & 41.1 & 61.0 \\ 
6 & 32.8 & 44.4 & 31.9 & 43.9 & 34.0 & 47.1 & 34.4 & 46.8 & 34.8 & 47.0 \\ 
7 & 35.7 & 44.6 & 36.7 & 44.2 & 34.8 & 45.1 & 35.9 & 45.9 & 34.3 & 47.9 \\ 
8 & 37.5 & 47.2 & 35.0 & 44.4 & 37.9 & 47.0 & 36.4 & 45.8 & 37.0 & 45.6 \\ 
9 & 55.9 & 59.9 & 55.8 & 61.3 & 55.6 & 61.2 & 56.9 & 61.6 & 57.0 & 62.5 \\ 
10 & 38.7 & 57.6 & 38.0 & 55.3 & 38.4 & 57.5 & 38.2 & 59.5 & 38.6 & 57.8 \\ 
11 & 43.4 & 44.6 & 44.8 & 45.8 & 43.1 & 45.1 & 44.5 & 46.0 & 42.8 & 43.0 \\ 
12 & 42.8 & 42.4 & 41.1 & 44.8 & 41.6 & 44.5 & 43.6 & 46.1 & 41.1 & 45.6 \\ 
13 & 0.1 & 84.5 & 0.4 & 85.0 & 0.3 & 84.7 & 0.2 & 85.8 & 0.2 & 86.2 \\ 
14 & 57.0 & 56.9 & 55.5 & 57.0 & 56.4 & 54.8 & 55.3 & 55.5 & 56.5 & 59.2 \\ 
\hline
\end{tabular}
\end{table}

\end{document}